\documentclass[runningheads]{llncs}
\usepackage{cite}
\usepackage{amsmath,amssymb,amsfonts}
\usepackage{algorithmic}
\usepackage{graphicx}
\usepackage{textcomp}
\usepackage{url}
\usepackage{xcolor}
\usepackage{color}
\usepackage{colortbl}
\usepackage{tabularx}
\usepackage{multirow}
\usepackage{subcaption}
\usepackage{todonotes}
\usepackage{paralist}

\definecolor{mygreen}{HTML}{00FF00}
\definecolor{myred}{HTML}{ffcc99}
\definecolor{myblue}{HTML}{9FC1FE}

\begin{document}
\title{When expertise gone missing: Uncovering the loss of prolific contributors in Wikipedia}
\titlerunning{When expertise gone missing}
%
\author{Paramita Das\inst{1} 
\and
Bhanu Prakash Reddy Guda\inst{2}
\and
Debajit Chakraborty\inst{1}
\and\\ 
Soumya Sarkar\inst{3}
\and
Animesh Mukherjee\inst{1}}
\authorrunning{P. Das et al.}
%
\institute{IIT Kharagpur, Kharagpur, India - 721302\\ \and
Adobe Research, Bangalore, India - 560087\\
\and
TU Darmstadt, Darmstadt, Germany - 64289\\
}

\maketitle              
\setcounter{footnote}{0}

\begin{abstract}
Success of planetary-scale online collaborative platforms such as Wikipedia is hinged on active and continued participation of its voluntary contributors. The phenomenal success of Wikipedia as a valued multilingual source of information is a testament to the possibilities of collective intelligence.
Specifically, the sustained and prudent contributions by the experienced prolific editors play a crucial role to operate the platform smoothly for decades. However, it has been brought to light that growth of Wikipedia is stagnating in terms of the number of editors that faces steady decline over time. This decreasing productivity and ever increasing attrition rate in both newcomer and experienced editors is a major concern for not only the future of this platform but also for several industry-scale information retrieval systems such as {\em Siri, Alexa} which depend on Wikipedia as knowledge store. In this paper, we have studied the ongoing crisis in which experienced and prolific editors withdraw. We performed extensive analysis of the editor activities and their language usage to identify features that can forecast {\em prolific Wikipedians}, who are at risk of ceasing voluntary services. To the best of our knowledge, this is the first work which proposes a scalable prediction pipeline, towards detecting the {\em prolific Wikipedians}, who might be at a risk of retiring from the platform and, thereby, can potentially enable moderators to launch appropriate incentive mechanisms to retain such `would-be missing' valued {\em Wikipedians}.

\end{abstract}
\section{Introduction} \label{sec:intro}
Wikipedia has emerged as an immensely popular digital encyclopedia, over the last decade. This is primarily attributed to the ``be bold'' policy, which permits anyone to contribute on almost all wikipages. Encouragement toward global collaboration 
generates hundreds of millions of views monthly which resembles Wikipedia as a reliable source of information, irrespective of socio-economic backgrounds and cultures~\cite{samoilenko2018don,lemmerich2019world}.
The workhorse behind this success story is a large pool of voluntary editors. These group of people maintain Wikipedia pages behind the scenes which includes creating new pages, changing contents, making sure that the fact provided is appropriate etc. It is imperative for the survival of this platform that this resource is nurtured so that they can operate with due diligence.  

In these lines there have been several works, which attempt to understand diverse roles editors play in the community~\cite{Zhang2010Expert,yang2016did, muric2019collaboration}. These studies revealed valuable insights about editors' motivation, group behaviour and productivity, thus further reinforcing our knowledge about community health. A complementary direction of exploration includes developing personalised recommendation system to identify appropriate pages and mentors for a newbie Wikipedia editor so that seamless onboarding to the platform can be achieved~\cite{balali2018newcomers,yazdanian2019eliciting}. However notwithstanding former efforts, there has been a steady attrition of Wikipedia editors~\cite{economist}. One possible reason could be personal disillusionment from the project. However, some notable reasons are \textit{increased bureaucracy} and \textit{incivility}~\cite{konieczny2018volunteer, ciampaglia2015moodbar}. In particular, to maintain the high quality standards of encyclopedic content, Wikipedia community has gradually become impermeable for the editors, resulting in unreasonable abuse of power such as blocking of users, deletion of good faith edits, and unexplained conflict arbitration~\cite{kiesel2017spatio}. Out-flux of contributors has been a chronic problem for other knowledge sharing platforms as well such as Yahoo! Answers, Baidu Knows etc.~\cite{dror2012churn,kairam2012life,danescu2013no}.

\noindent\textbf{Loss of Wikipedians is a loss of Wikipedia}: Editors, or Wikipedians~\cite{Panciera2009Power, sarasua2019evolution} are the soul of Wikipedia; without their active participation, Wikipedia will eventually become stagnant. As mentioned earlier, the loss of editors and simultaneously \textit{editor retention}~\cite{gallus2017fostering} is a Wikipedia-wide problem. 

\noindent\textit{Related wikiprojects}: Several wikiprojects~\cite{teahouse}, hosted by Wikipedia aim to construct a stronger bonding among Wikipedians and also figuring out several measures~\cite{wpAct,wpSurvive} to establish the objective of editor retention.  
There exists corresponding wikiproject~\cite{wpRetention} dedicated to find the cause of editor depletion and early identification of editors at risk of forsaking Wikipedia. Often expert editors withdraw themselves because of the discontent with Wikipedia's policies and widespread norms resulting them into semi-retired or permanent-retired often not disclosing the prime reasons behind the retirement. As an example, the Wikipedia project (WP:MISS)\footnote{\url{https://en.wikipedia.org/wiki/Wikipedia:Missing_Wikipedians}} maintains a list of experienced Wikipedians who have made no edits to any article after a fixed time point in their life and are defined as {\em missing Wikipedians}. All of these editors contributed at least 1000 edits in their lifetime and can be considered as \textit{prolific editors}. Even after being inactive, many of them are still featured among top editors\footnote{\url{https://tinyurl.com/3ewzb468}}. As of June 2020, this list includes 1226 such editors whom we refer to henceforth as \textit{missing Wikipedians} or \textit{missing editors} interchangeably. In contrast, the pool of editors who are continuing to contribute as active participants till date, will be referred as \textit{active editors} hereafter.  Recently, a wikiproject named as \textit{Community Health Metrics: Understanding Editor Drop-off}\footnote{\url{https://tinyurl.com/3e6jj6zz}} has been initiated to understand the editor drop-off across multiple language versions of Wikipedia. The project aims to understand various dynamics in editors' life-cycle with a special attention to the veteran editors of the community who are at the risk of leaving Wikipedia. Our work is in line with the objectives of these projects - identifying the decline of prolific editors whose contributions are measured significant in the community. We observed that majority of editors faced some form of impediment from the Wikipedia community that curbed their interest in further editing. Finally, these missing editors have left the platform and there is no assurance of their return\footnote{\url{https://tinyurl.com/35n624a6}}. Most of these editors do not reveal the explicit reason for leaving Wikipedia. Some editors have pointed out typical instances of disagreement, bureaucracy on their personal user talk pages that influenced them for announcing retirement. Among many example reasons that we observed on editors' user pages, a typical example is the following-

\noindent\textit{``I have left Wikipedia. I do not see it as acceptable to have advertisements, whether they be for brand identity or for a product, on Wikipedia.'' --\textbf{Im*****v}}

\noindent The above text shows an editor's disagreement with the community for compromising NPOV policy\footnote{\url{https://en.wikipedia.org/wiki/Wikipedia:Neutral_point_of_view}} of Wikipedia as a prime cause of his/her withdrawal from Wikipedia.

\noindent\textbf{Editor attributes as a cue?}: Our hypothesis is that the early signals of retirement of a missing editor could be hidden in his/her last trail of \textit{editing activities}. In other words, the editing activity patterns of the missing editors would have difference with the existing active editors. Further, the typical sentiments, emotions expressed on user talk pages should strike different attitude of missing editors compared to the active group of editors.
Due to the challenges and hindrance of the platform, missing editors might follow different profile of quality as compared to active editors. Based on this hypothesis, in this paper, we develop a framework for the discovery of key editors who have stopped content generation in Wikipedia. For this purpose, we first identify 1146 editors who have no edits in the calendar year of 2020 and denote them as missing editors. We next select a set of 2569 editors whose editing activity (i) are ongoing and (ii) match the overall editing activity level of the missing editors. This second set is denoted as active editors. We compare the two sets of editors on a longitudinal scale in the attempt to identify various properties related to their activity patterns and language usage that could significantly differentiate them. Our main research questions are as follows.

\noindent\textit{RQ1:} To what extent activity/linguistic/quality features of editors can help us discern missing from active editors?

\noindent\textit{RQ2:} Using the discriminative editor activities can we predict currently prolific editors who have a possibility to leave the platform in the near future?

\noindent\textbf{Our contributions}: We make the following novel contributions in this paper.
\begin{compactitem}
    \item We curate a first ever dataset of missing editors, a comparable dataset of active editors along with all the associated metadata that can appropriately characterise the editors from each dataset. (see section~\ref{sec:dataset})
    \item First we put forward a number of features describing the editors (activity and behaviour) which portray significant differences between the active and the missing editors. (see section~\ref{sec:features})
    \item Next we use SOTA machine learning approaches to predict the currently prolific editors who are at the risk of leaving the platform in near future. Our best models achieve an overall accuracy of 82\% in the prediction task. (see section~\ref{sec:classification})
    \item We perform rigorous ablation studies to provide further insights into our results. We discuss various nuanced observations that get manifested in the course of our study. An intriguing finding is that some very simple factors like how often an editor's edits are reverted or how often an editor is assigned administrative tasks could be monitored by the moderators to determine whether an editor is about to leave the platform. (see section~\ref{sec:classification})
\end{compactitem}

To the best of our knowledge this is the first work which proposes an automatic approach to predict, early on, the editors who have a propensity to leave the platform soon. We believe that the moderators can use these attributes to launch suitable \textit{platform governance}~\cite{gorwa2019platform} measures to appropriately incentivize these editors and, thereby, retain them on the platform. Further the observations and insights that we obtain from our results can he useful in designing the specific incentive strategies~\cite{aaltonen2015building}.

\section{Related work}
The reason behind the popularity of Wikipedia is often attributed to the hundreds of thousands of volunteers from all around the world, but several tens of thousands of Wikipedians and their collaboration are very crucial for the generation and maintenance of healthy and informative content~\cite{suzuki2012mutual,proffitt2018leveraging,kittur2008harnessing}. Researchers are trying to overcome different challenges disseminated by open source of knowledge and information especially the quality and trust issues~\cite{pinto2018towards,chen2019weakly} of this gigantic platform.

With the constant effort of regulation and maintenance of encyclopedia content, it appears that Wikipedia is at the peak of its popularity; however, experts noted that it is at the danger of sharp decline of its active editors~\cite{halfaker2013rise,zhang2012leave}. The revert of edits is highly effective in controlling vandalism~\cite{green2017spam}, sometimes it becomes harsh to the editors, especially the newly joining ones~\cite{halfaker2011don}. The topic of volunteer retention~\cite{konieczny2018volunteer} is one of the basic concerns of all peer-production systems, tied to their preliminary survival. Hence, researchers of Wikipedia shed light on the \textit{retention} of \textit{new} contributors~\cite{choi2010socialization,morgan2018evaluating,li2020successful}. However, works on early identification and retention of \textit{prolific editors} at risk of retiring are scarce. Therefore, in this work we set out to develop an automatic framework that can accurately identify editing activity signatures that point to the early signals resulting in the abdication of an editor.

\section{Dataset description} \label{sec:dataset}
Wikipedians are the prime elements of our experiment. In order to curate  two sets of editors -- missing and active, we follow a two step approach.

\noindent\textbf{List of missing editors}:
First, we prepare a list of editors who are mentioned as \textit{missing} by Wikipedia. We extracted the list of 
missing Wikipedians\footnote{\url{https://en.wikipedia.org/wiki/Wikipedia:Missing_Wikipedians}} and their last date of contribution using XTools\footnote{\url{https://www.mediawiki.org/wiki/XTools/API}}. Wikipedia has mentioned 1226 such editors in the list at the time of our dataset collection (June, 2020). From this list we remove those that we found still contributing in 2020 despite being declared as missing by Wikipedia. This results in 1146 unique editors who further have not edited in the calendar year 2020.

\noindent\textbf{List of active editors}: Next, we collect a set of 2569 editors who are still active, i.e., are still editing different Wikipedia articles. In order to build a comparable set of active editors, we performed the following. 
\begin{compactitem}
        \item First, we collect the list of top 60 pages (top 20 from each namespace -- `Main', `Wikipedia' and `Talk') that have been most frequently edited by each missing editor. This leads us to an initial set of 68,760 Wikipedia pages. 
    \item Next we look into the 100 latest revisions of each of the 68,760 pages and the editors who contributed to the latest revisions are extracted from the revision history of the corresponding pages. Further, for every missing editor, only top 10 active editors are selected based on the frequency of their edits on the above mentioned top edited 60 pages. Formally, for a missing editor $M$, the editors $A_{1},A_{2},...,A_{10}$ are assumed to be the most active if they have the highest edit counts in the last 100 revisions of the top edited pages $P_{1}, P_{2},P_{3},...,P_{60}$ by the editor $M$. These editors are placed into the pool of active editors. For every missing editor we perform this exercise leading us to a set of 5213 unique active editors. We confirm that none of these editors are listed in the missing Wikipedians page.
    \item As a final step, we compute the average number of edits per day for the active editors and the missing editors. We next compute the mean of the average edits per day for the missing editors (say $m$). Next we compute the L2-norm of $m$ with the average number of edits per day of each of the active editors. We consider those active editors for whom the L2-norm is within one standard deviation of $m$. This results in 2569 active editors. The difference in the distributions of the average edits per day for the missing and active editors so chosen is not statistically significant (Mann Whitney U test, $p=0.14$) thus making them comparable.
\end{compactitem}

\section{Feature space design} \label{sec:features}
In this section we describe our feature space which we develop to differentiate the active from the missing editors. The features can be categorised into three broad classes--

\noindent Activity features, Quality features, Linguistic features. 

Resources for the work is available at: \url{https://github.com/paramita08/ Missing-Active-Wikipedians}

\subsection{Activity features}
The features we propose in this category describe the activity patterns of editors. The choice of these features is motivated by the fact that these would reflect different aspects of editor's interactions with the community over time. These features, we believe should be useful to capture the temporal change in the behavior of the editors who have a possibility to leave the platform. We extract the details of the latest 50 non-automated edits of every user using the \textit{User API} of XTools\footnote{\url{https://www.mediawiki.org/wiki/XTools/API/User#Non-automated_edits}} and identify the following activity features. 

\begin{figure}[h]
\begin{subfigure}[b]{0.46\textwidth}
  \centering
  \includegraphics[width=\textwidth, height=40mm]{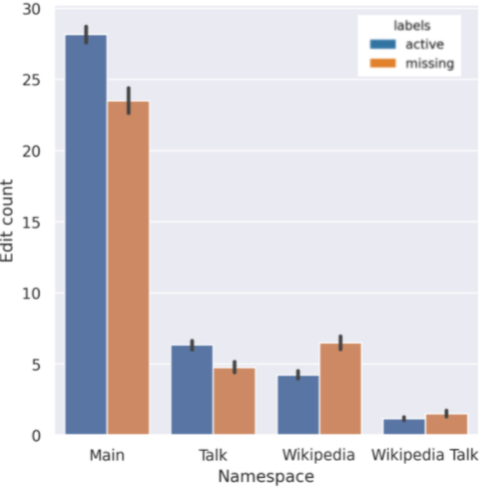}
  \label{feat_namespace}
  \caption{\label{namespace}The average count of edits in different namespaces of active and missing editors.}
\end{subfigure}
\hspace{0.2cm}
\begin{subfigure}[b]{0.46\textwidth}
  \centering
  \includegraphics[width=\textwidth, height=40mm]{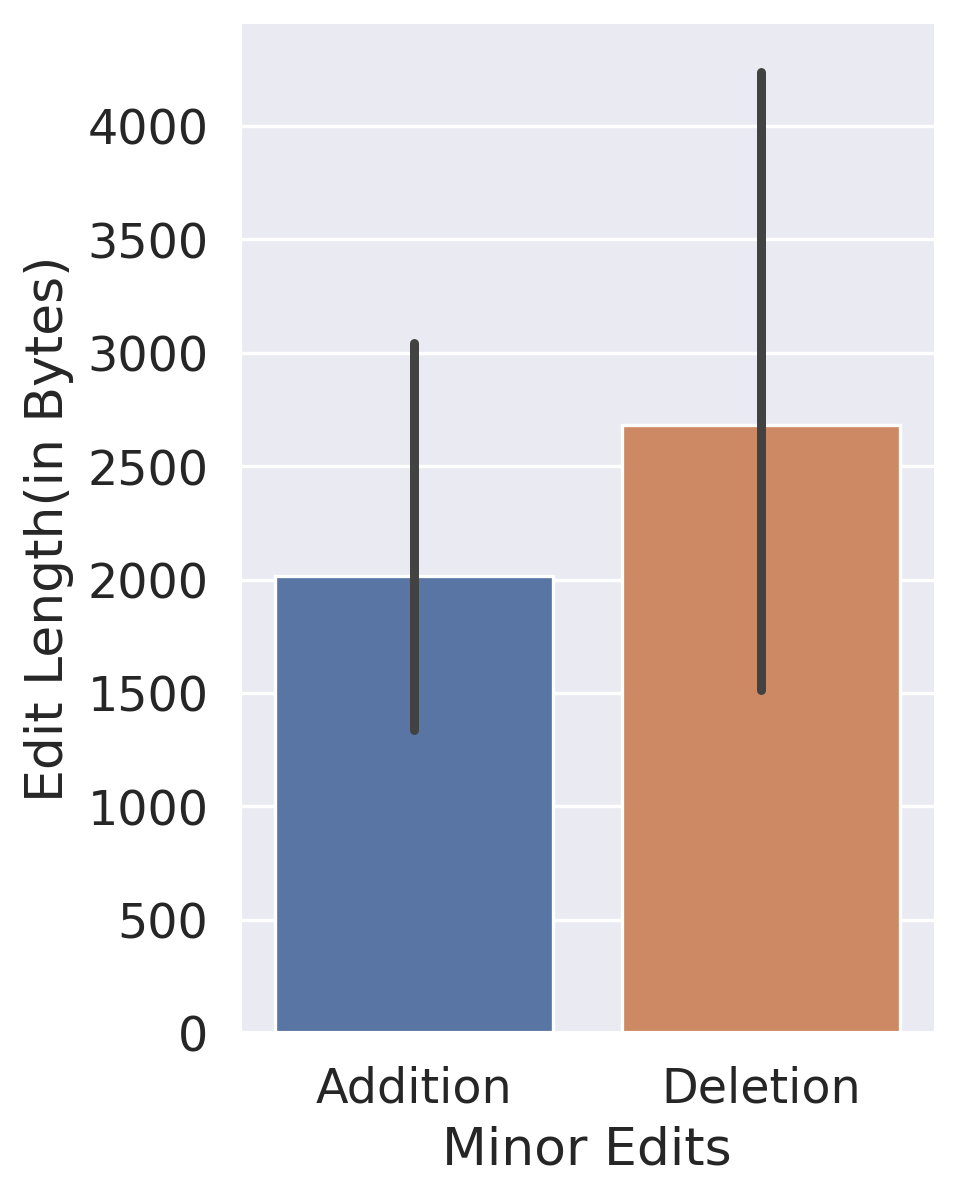}
  \label{feat_minor}
  \caption{\label{edits_bytes}Average length of edits(in bytes) in minor edits (addition and deletion) by missing editors.}
\end{subfigure}
\caption{Plots showing different feature values for the two classes of editors.}
\label{fig:feat_analysis}
\end{figure}

\noindent\textit{Edits in different namespaces}: Among different namespaces\footnote{\url{https://en.wikipedia.org/wiki/Wikipedia:Namespace}} mentioned, we consider the editor contribution in 4 namespaces - article pages i.e., namespace 0; article talk pages, i.e., namespace 1; administrative main pages - Wikipedia, i.e., namespace 4 and the administrative talk pages - Wikipedia talk, i.e., namespace 5. Majority of contribution of editors is restricted in these respective namespaces~\cite{maki2017roles} and we denote the four namespaces as F1, F2, F3 and F4 following the order as mentioned above. Our hypothesis is that the frequency of edits in the first two namespaces (i.e., 0, 1) will be affected drastically for the missing editors in their latest series of contributions. We have included these two namespaces to observe if the missing editors exhibit a decline in participation in Wikipedia's policy decisions. As shown in Figure \ref{fig:feat_analysis} (a), the average count of edits in all the namespaces except namespace 5 is higher in case of active editors. This trend implies a slow decline in contribution from missing editors possibly pointing to their absence in near future.

\begin{figure} [h]
    \centering
    \includegraphics[width=\textwidth,height=3.5cm]{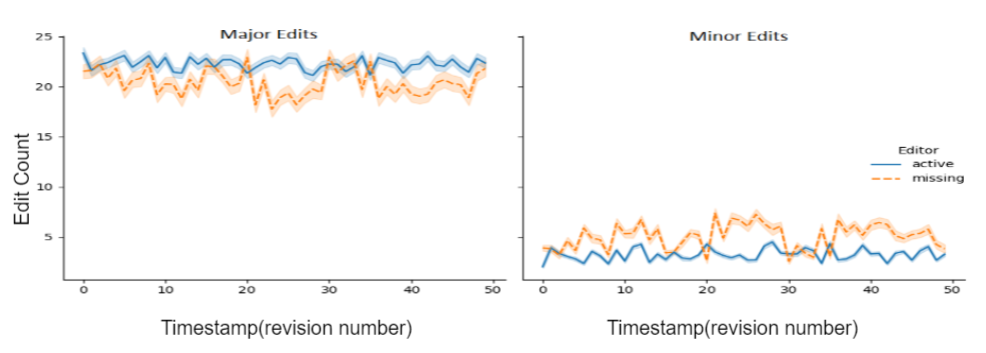}
    \caption{Temporal trends showing major and minor edit counts individually over the latest revisions for both the sets of editors. The cumulative count at every revision exhibits that missing editors contributed more minor edits than the active editors and vice-versa. }
    \label{fig:temporal_edit}
\end{figure}

\noindent\textit{Major and minor edits}:
A \textit{minor}\footnote{\url{https://en.wikipedia.org/wiki/Help:Minor_edit}} edit is the one denoting minor changes such as typographical errors, formatting errors, reversion of definite vandalism etc. that the editors believe need no further reviews. In contrast, any contribution that changes the article content and needs to be reviewed for the acceptability to the community of editors is denoted as \textit{major} edit. We hypothesize that an increase in the number of minor edits as compared to major edits can bear an early signature of their disengagement, finally abdicating the platform. Hence, we included the count of edits in both categories - major (F5) and minor (F6) of the latest revisions in our feature set. The cumulative edit count of major and minor edits in Figure \ref{fig:temporal_edit} show that for major edits, active editors exhibit a steady state in the count over time but the decline is extremely unstable for the missing editors and almost always below that of the active editors. On the other hand, close to their retirement, the missing editors seem to be engaging themselves in an increasingly more number of minor edits as compared to the active editors.

\noindent\textit{Length of the edits}: The average length of the edits is captured by this feature in terms of the bytes added or deleted per edit. Further, we classify this feature into four categories -
(i) addition in major edits (F7), (ii) deletion in major edits (F8), (iii) addition in minor edits (F9), (iv) deletion in minor edits (F10). In each of the cases, we compute the average bytes added (or deleted) in case of a major (or minor) edit over the latest contribution of the editors. Once again the hypothesis is that the length of the edits done by the missing editors shall diminish over time.  An interesting observation that we have here is that the missing editors seem to engage more in deleting minor edits than adding minor edits (see Figure \ref{fig:feat_analysis}(b)).  

\noindent\textit{Span (in months)}: We compute the time span (denoted as F11) in months taken by each editor to complete the latest revisions of contribution. The span is expected to be larger in case of missing editors compared to the active editors indicating a possible loss of overall interest in the platform. We find that the average span in case missing editors is 2.04 months while for active editors this is 0.82 months.

\noindent\textit{ORES score of the edits}: ORES\footnote{\url{https://www.mediawiki.org/wiki/ORES}}, the web-service API by Wikimedia Foundation is used in automating several wiki management tasks such as assigning scores to individual edits of editors and predicting the edits to be damaging or good-faith. We compute the average ORES score (represented as F12) for each of the editors in our dataset over their contribution span. In addition, we consider the count of the good faith (denoted as F13) vs the damaging (denoted as F14) edits for every editor in our dataset. On average we did not see any difference in the quality of edits done by the two classes of editors. Further both the classes have a very small number of damaging edit contributions.

\subsection{Quality features}
Although the content of any Wikipedia article is factual, its quality heavily depends on the dexterity of the editors. We formulate the following quality features that could be potential indicators of the future churn.

\noindent\textit{Count of reverts}: Reverting is the way of reversing a prior edit or undoing the effects of one or more edits, which typically results the article to be restored to a version that existed sometime previously. Reverts are discouraging to any editor, irrespective of their experience~\cite{kiesel2017spatio,halfaker2011don} and can potentially impede an editor toward making future contributions to the platform. To investigate this, we scrape the number of reverted edits of every individual editor in their respective top 50 most edited pages in the main namespace only. This is since almost all editors make their largest chunk of contributions to the main namespace. From
Table~\ref{tab:quality_ft} we observe that missing editors have experienced a larger number of reverts which is in line with our hypothesis that more reverts potentially disengage editors from the platform. Further we revisit a few nuanced cases in which reverts and the comments with the reverted edits had a negative impact on the missing editors. As shown in the Table~\ref{tab:revert_comments}, a missing editor received several foul comments with the reverted edits which were posted on her talk page publicly. Finally, she stopped editing after experiencing a lot of reverts toward the end of 2004 with a clear message declaring her retirement on the user page in 2005. We also look into the talk pages of active editors and find that they had experienced less sensitive comments for their reverted edits and an example scenario is depicted in the second row of the Table~\ref{tab:revert_comments}. 

\begin{table}[h]
    \centering
    \begin{tabular}{|c|c|c|}
        \hline
        Quality & Missing $(\mu, \sigma)$ & Active $(\mu, \sigma)$\\
        \hline
        Avg. revert count per page & (0.031, 0.041) & (0.026, 0.031) \\
        \hline
        Admin score & (734.38, 165.38) & (823.86, 174.93)\\
        \hline
    \end{tabular}
    \vspace{1em}
    \caption{Mean ($\mu$) and std. deviation ($\sigma$) of the quality features for missing and active editors.}
    \label{tab:quality_ft}
\end{table}

\noindent\textit{Admin score}: Wikipedia maintains a pool of editors, known as admins, who are responsible to perform various administrative tasks and the tasks are pivotal in monitoring various quality issues of Wikipedia. These set of people act voluntarily and every wikiproject tries to find the admins from among the prolific and  experienced editors. XTools provides the \textit{admin score}\footnote{\url{https://xtools.readthedocs.io/en/3.1.6/tools/adminscore.html}}, intended to find how suitable an user is for serving as an admin. The more the admin-score, the higher is the chance of the editor to become an admin. We observe in Table \ref{tab:quality_ft} that active editors on average have a higher admin score and therefore higher chances of being an admin as compared to the missing editors. 

\begin{table*}[h] 
    \centering
    \begin{tabular}{l||p{80mm}}
        \textbf{comments received by} & \textbf{comments} \\
        \hline
         a missing editor & \textit{if that happens I will, with considerable regret, withdraw from Wikipedia altogether. An encyclopedia that can't or won't defend itself against Stalinist and LaRouche wreckers will never succeed and doesn't desrve to.}  \textbf{--- A****, 01:05, 16 Nov 2004 (UTC)}\\
         \hline
         \hline
         an active editor & \textit{Greetings. I noticed you had undid my edits on H**** C**** and J**** K****, specifically the ones....I've since reverted them. But I'm opening discussion here, in case there's disagreement. Looking forward to your thoughts.} \textbf{--- Ga****, 14:06, 23 Dec 2015 (UTC)}\\
         \hline
    \end{tabular}
    \vspace{1em}
    \caption{Example comments on editors' talk page showing comments received with reverted edits. The comments received by an active editor is less harsh as compared to a missing editor.}
    \label{tab:revert_comments}
\end{table*}

\subsection{Linguistic features}
As we have discussed earlier a number of editors have shared their basic information, Wikipedia-related activities, awards and badges earned etc. and sometimes their grievances\footnote{\url{https://en.wikipedia.org/wiki/User:Bcrowell}} about the platform on their user pages that finally hold them back in continuing further contributions. We assume that these pages to be the profile of the editors and leverage the user generated text in characterizing the two sets of editors. We extract the text from the HTML version of the latest editor page and perform pre-processing to remove various links and HTML tags. We next extract various features from this pre-processed text.

\noindent\textit{POS tags}: We compute the POS tags of the text in the editor pages using the NLTK parser~\cite{loper2004nltk}. Prior to tagging, we tokenized the sentences and words common to both the groups of editors (missing and active) are removed from the corpus. We also remove the stop words and the non-English words. We observe several frequent POS tag categories such as ``JJ'', ``NN'', ``NNS'', ``VBD'', ``VBP'' that exhibit significant differences between the missing and the active editors.

\noindent\textit{Empath categories}: Empath is a text analysis tool~\cite{fast2016empath} which can identify psycho-linguistic signs hidden in a text in the form of pre-validated lexical categories. 
For our purpose, we compute the normalised value for each predefined category for each sentence. We then average out this value for all the sentences present in the profile of an editor. We observe that a number of lexical categories (21 in number), i.e, ``Internet'', ``Noise'', ``Trust'', ``Reading'', ``Violence'', ``Negative Emotion'', ``Positive Emotion'' etc. show the difference in distributions for two classes of editors is statistically significant as per the Mann-Whitney U test. However the absolute differences between the two classes in terms of the values of the various lexical categories is negligibly small. However the absolute differences between the two classes in terms of the values of the various lexical categories is negligibly small.

\begin{figure} [h!]
    \centering
    \includegraphics[width=0.8\columnwidth, height=40mm]{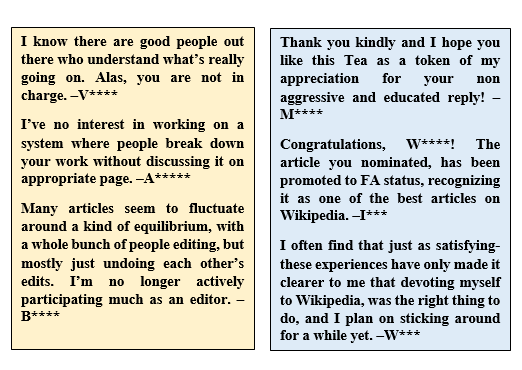}
    \caption{A set of sample sentences, narrated by some anonymous editors belonging to the two prominent sentence clusters (see the t-SNE plot in Figure~\ref{subfig:tsne}). In the cluster of the missing editors, we find sentences expressing grudges against the platform (\colorbox{myred}{red} sentences). In the cluster of active editors, we find positive sentences about the platform (\colorbox{myblue}{blue} sentences).}
    \label{fig:sentences}
\end{figure}

\noindent\textit{Sentence vector}: We generate the sentence vector using the Universal Sentence Encoder \cite{conneau2017} which outputs a 512 dimensional vector representation of the text. We compute the sentence vector for every editor which can be thought of as a summary of the language usage of the users on the platform. To visualise how different the sentence vectors from the two classes are, we plot them using t-SNE \cite{van2008visualizing}. We consider the first two principal components of the vectors for visualisation. Figure \ref{subfig:tsne} shows that there is no clear separation between the sentence vector drawn from the two classes. However, there are at least two small but very distinct pockets of vectors from a particular class clustering together expressing opposing sentiments toward the platform. We  handpick a few examples from the talk pages of editors belonging to each of these two clusters. These are shown in Figure \ref{fig:sentences}. The difference in sentiments toward the platform is apparent in the selected sentences from the two clusters. While missing editors seem to express grudges against the platform, the active editors mostly narrate their positive experiences with the platform.

\subsection{Feature correlation}
\begin{figure} [h]
\centering
  \begin{subfigure}[b]{0.48\linewidth}
  \centering
    \includegraphics[width=\linewidth,height=30mm]{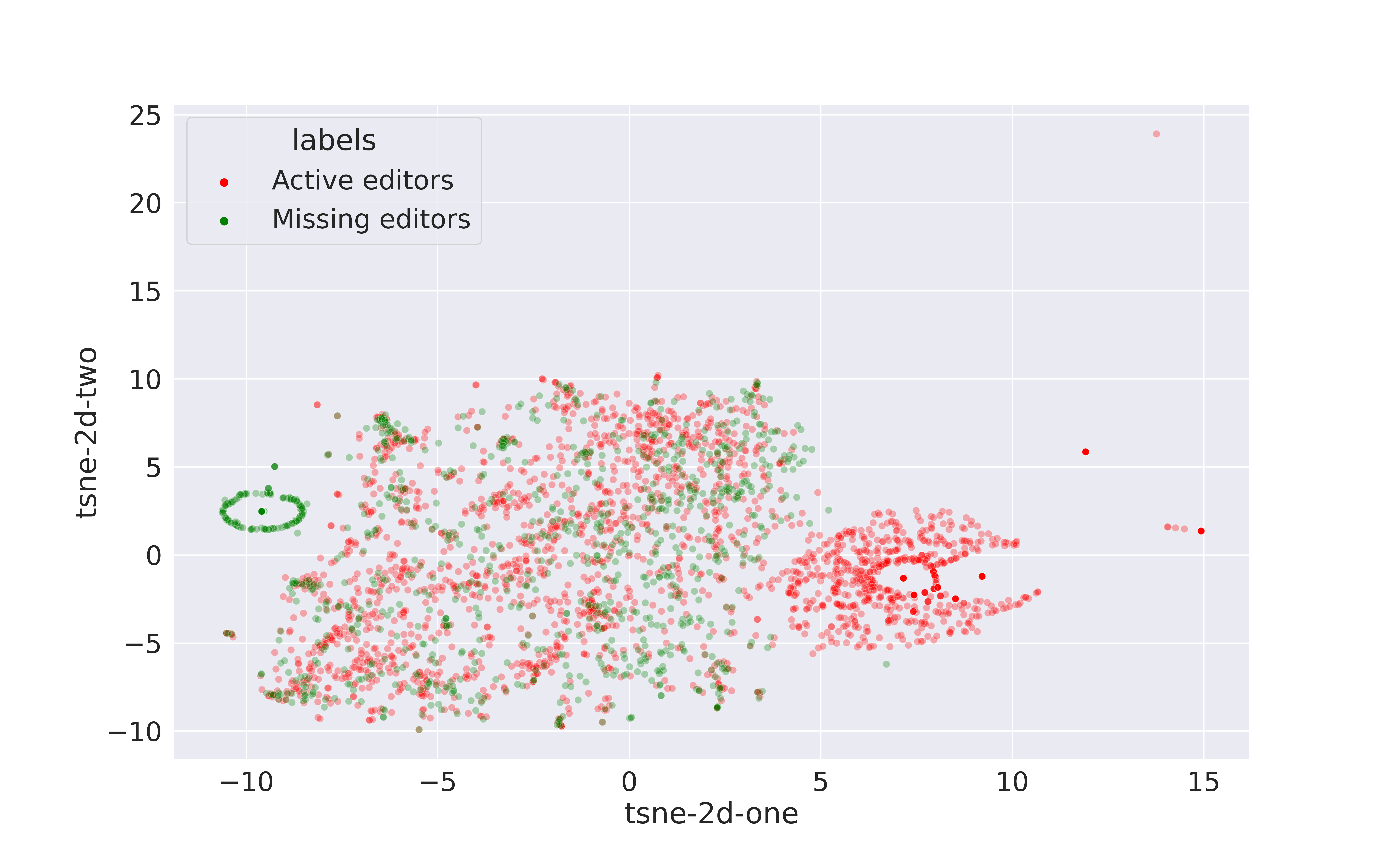} 
    \caption{t-SNE visualization}
    \label{subfig:tsne}
  \end{subfigure}
  \hspace{0.2cm}
  \begin{subfigure}[b]{0.48\linewidth}
  \centering
    \includegraphics[width=\linewidth,height=30mm]{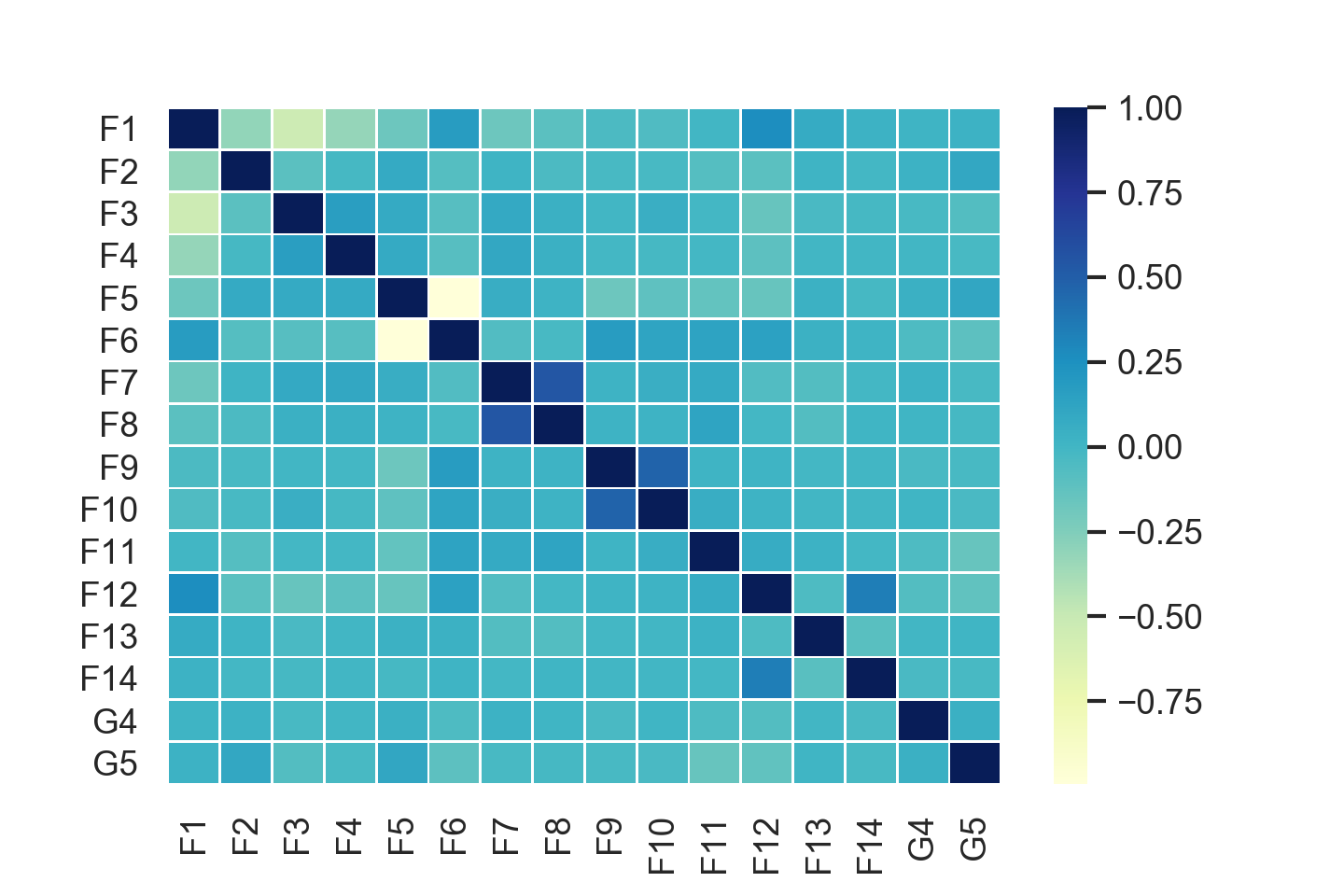} 
    \caption{Heatmap visualization}
    \label{subfig:correlation}
  \end{subfigure}
  \caption{(a) t-SNE showing sentence vectors expressed by two sets of editors-- missing (in green) and active (in red) and (b) Heatmap showing the correlation between the various features from the activity and quality categories.}
\end{figure}

To examine the relationship among the different features described above, we compute the Pearson's correlation coefficients~\cite{benesty2009noise} between the all-pair features. Figure \ref{subfig:correlation} shows the correlation coefficient matrix for the 16 features (except the linguistic features). We observe that some features are positively correlated and some features are negatively correlated (e.g., major edits and minor edits) with each other. However almost all the coefficients are in the range -0.05 to 0.05 (i.e., $\sim0$ correlation) except for the pairs (addition in major edits, deletion in major edits) and (addition in minor edits, deletion in minor edits), which belong to the same category. We do not include the linguistic features in this study as they are characteristically very different from these features.

\section{Classification of editors} \label{sec:classification}
In this section we use the mentioned features to automatically classify the missing editors from the active ones. However, in order to make the model outcomes interpretable we incrementally include the features to better understand their effect on the overall performance. We name the feature groups as follows-
\begin{compactitem}
    \item Activity features as G1
    \item POS tags and Empath  as G2
    \item Sentence vector as G3
    \item Admin score as G4
    \item Count of reverts as G5
\end{compactitem}

\noindent\textit{Classification model}: Corresponding to each editor in our dataset, we construct a normalised feature vector. We experiment with multiple classifiers such as Random Forest, Logistic Regression SVMs, XGBoost, AdaBoost. In this task, a binary classifier is built to predict if the given editor is a missing or active one. The train-test spilt for our experiments is set to 80:20. We use accuracy along with weighted precision, recall and F1-score as the evaluation measures.

\noindent\textit{Results}: Among the different features, we observe that the 14 activity features, reverts and admin score seem to be performing much better than the linguistic features (see Table \ref{tab:classification_results}). The \textsc{G1} feature group, i.e., the activity features seem to be the strongest individual discriminator resulting in an accuracy of 75\% (F1-score 74\%). Together with \textsc{G4}, this accuracy goes to 78\% (F1-score 79\%).  Among the linguistic features, the sentence vector seems to be the most effective in conjunction with activity and the admin score features. Together this feature group \textsc{G1}, \textsc{G3} and \textsc{G4} attains an accuracy of 82\% (F1-score 81\%). The feature group \textsc{G1}, \textsc{G4} and \textsc{G5} also attains a similar accuracy of 82\% (F1-score 81\%). Larger groups of features do not seem to improve performance. These results further corroborate that missing editors hardly leave any special linguistic cues before they abdicate the platform.  

\begin{table*}[h]
\centering\scriptsize
\scalebox{1.0}{
    \begin{tabular}{|c|c|c|c|c|c|}
        \hline
        Features & Classifier & Precision & Recall & F-score &  Accuracy \\
        \hline
        \textsc{G1} & XGBoost & 0.74 & 0.74 & 0.74 & 0.75 \\
        \hline
        \textsc{G2} & XGBoost & 0.63 & 0.68 & 0.64 & 0.68 \\
        \hline
        \textsc{G3} & AdaBoost & 0.63 & 0.67 & 0.63 & 0.67 \\
        \hline
        \textsc{G1} $\oplus$ \textsc{G2} & Random Forest & 0.77 & 0.78 & 0.76 & 0.78 \\
        \hline
        \textsc{G1} $\oplus$ \textsc{G3} & XGBoost & 0.75 & 0.76 & 0.75 & 0.76 \\
        \hline
        \textsc{G1} $\oplus$ \textsc{G4} & AdaBoost & 0.78 & 0.79 & 0.79 & 0.78 \\
        \hline
        \textsc{G1} $\oplus$ \textsc{G5} & AdaBoost & 0.78 & 0.79 & 0.79 & 0.77 \\
        \hline
        \textsc{G1} $\oplus$ \textsc{G2} $\oplus$ \textsc{G4}& XGBoost & 0.77 & 0.78 & 0.77 & 0.78 \\ 
        \hline
         \rowcolor{green!30}\textsc{G1} $\oplus$ \textsc{G3} $\oplus$ \textsc{G4}& XGBoost & 0.81 & 0.82 & 0.81 & 0.82 \\
        \hline
        \rowcolor{green!30}\textsc{G1} $\oplus$ \textsc{G4} $\oplus$ \textsc{G5}& AdaBoost & 0.81 & 0.82 & 0.81 & 0.82 \\ 
        \hline
        \textsc{G1} $\oplus$ \textsc{G3} $\oplus$ \textsc{G5}& XGBoost & 0.78 & 0.79 & 0.78 & 0.79 \\ 
        \hline
        \rowcolor{green!30} \textsc{G1} $\oplus$ \textsc{G2} $\oplus$ \textsc{G4} $\oplus$ \textsc{G5}& Random Forest & 0.82 & 0.82 & 0.82 & 0.81 \\ 
        \hline
        \rowcolor{green!30} \textsc{G1} $\oplus$ \textsc{G3} $\oplus$ \textsc{G4} $\oplus$ \textsc{G5}& XGBoost & 0.80 & 0.81 & 0.80 & 0.81 \\
    \end{tabular}
    }
    \vspace{1em}
    \caption{Prediction outcomes: combination of different features. The green rows indicate the best result using the respective feature combination.}
    \label{tab:classification_results}
\end{table*}

\noindent\textit{Feature importance analysis}:
In this section we investigate the importance of the individual features in this feature group that consists of only activity and quality features (16 in all). To this purpose, we use three different ways to compute the importance.

\textit{Feature importance function}: 
We use the {\em Gini impurity} bases feature importance which is a standard technique in tree based ensemble methods~\cite{nembrini2018revival} to compute the feature importance values. We observe the features that dominate the top of the rank list are - reverts, ORES score, admin score, deletion in major edits, edits in namespace 0 and span.

\textit{Permutation importance function}: 
We compute feature importance using {\em permutation importance}~\cite{altmann2010permutation} and observe almost similar importance order as that of the previous function. The features that come at the top 6 places of the importance list are - reverts, ORES score, admin score, span, edits in namespace 0 and edits in namespace 4.

\textit{LIME}: LIME~\cite{ribeiro2016should} is an explainability tool meant to generate explanations regarding the workings of a ML model. We use all the instances in the test data for generating the explanations. We note the top five features returned by LIME that are responsible for the prediction outcomes for each instance. We observe that across all the instances the set of features $\{$reverts, admin score, edits in namespace 0, edits in namespace 1, ORES damaging \textit{true}, edits in namespace 4$\}$ present the best explanations.

Overall, we observe that the quality features are the most discrimiantory among all. This points to the fact that they constitute the best signals for identifying whether an editor is about to leave the platform. Platform moderators can deploy simple monitoring schemes to understand these signals early on and take preventive steps to minimise the loss of such prolific contributors.

\section{Discussion}
The primary objective of this work has been to bring forth the issue of the growing depletion of editors, especially the experienced editors in Wikipedia. One of the most important findings of our work is that missing editors can be differentiated from the existing active ones by a set of simple and interpretable characteristics of editors extracted from their interaction with the platform.  

\begin{figure}[h!]
    \centering
    \includegraphics[height=40mm, width=0.7\textwidth]{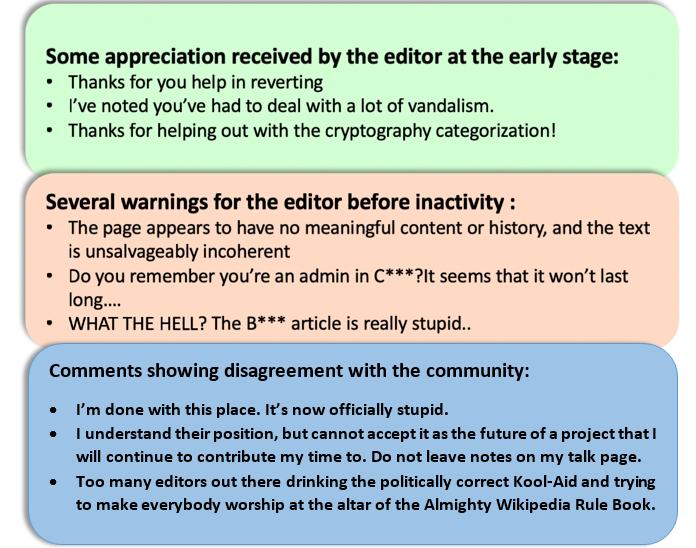}
    \caption{Examples showing talk page arguments of a missing editor. The text in \colorbox{mygreen}{green} colour box shows some appreciation received by the editor at early stage. In later phase, a number of toxic comments (shown in \colorbox{myred}{red} colour box) alleged  editing activity. A few examples from his user page (shown in \colorbox{myblue}{blue} colour box) indicates editor's dissatisfaction toward the community.}
    \label{fig:talk_convers}
\end{figure}

\noindent\textit{Dissatisfaction of prolific editors}: Further to test the accuracy of our model, we have observed a number of specific cases where the classifiers are confident in predicting the missing editors. We revisited their talk pages and found that in many such cases the editors had to face bullying\footnote{\url{https://en.wikipedia.org/wiki/Wikipedia:WikiBullying}} from other editors. Those bullying instances are usually mentioned as a part of the article talk pages. Sometimes the user has also been found to express his/her grievances on their personal user page. Figure \ref{fig:talk_convers} shows few example cases of toxic arguments and explicit reasons for leaving the platform that the editor had to face before they abdicated. Although a number of editors have stopped their contribution silently, the external causes of discontent with the platform that trigger the churn of prolific editors need to be considered carefully. 

\noindent\textit{Enabling platform moderation}: We collected a total of 8 false positive cases (mentioned as active but predicted as missing) in which our model is highly confident (confidence probability $> 0.8$). We found that out of these sample cases, the last edit for 4 editors were done latest in $2017$. We believe that all these editors should be potentially included in the list of missing editors but somehow got overlooked by the Wikipedia community. In fact, one out of these 4 cases have been actually identified as a missing editor in the most recent edition of the missing editor list, i.e., June 2021. From an application point of view, our system can monitor the current editor pool and flag any missing editor which the community may overlook. Suitable policies may be adopted to retain/reinstate such editors.

\noindent\textit{Design implication}:
Our idea was to keep the model as much interpretable as possible so that it is easy to understand whether the editing activity and the language usage across the platform are useful for classification. Further, our dataset is unbalanced and the total number of data points is limited to only $3715$. Hence, in our work, we have followed simple feature based classification model. Moreover, we proposed a novel task in which the feature set design is such that they can be calculated in all category of Wikipedia articles. 
The existing literature tried to divide editors either with respect to a certain number of Wikiprojects or based on a set of Wikipedia articles. Therefore suitable baselines are scarce. Thus here we consider feature ablations as baselines as is normative in many NLP applications~\cite{mowery2017feature} where suitable baselines are not available .

\section{Conclusion}
In this paper we take a deep dive into the emerging issue of editor depletion in Wikipedia. Our investigation shows that longitudinal activity traces as well as linguistic clues in their profile, i.e., user talk pages provide necessary signals to ascertain whether a prolific contributor will be quitting the Wikipedia platform in near future. We identify a number of important features that summarise the level of activity and behaviour of an editor and can be used to investigate the casual connection for leaving the platform. For instance, if two missing editors have almost equivalent qualities and extremely similar editing activity patterns and have retired almost at the same time, then it might well be the case that both of them left the platform for very similar reasons. In future, we plan to use these features to identify the causes of a group of editor leaving Wikipedia. This would allow for a better streamlining of the governance measures based on the nature of the cause of disengagement. Upon acceptance, we plan to place all our data and code in the public domain to facilitate further research in this area.

%
%
%
\bibliographystyle{splncs04}
\bibliography{ref}

\end{document}